\documentclass[reprint,amsmath,aps,prb,superscriptaddress,citeautoscript]{revtex4-2}
\usepackage[utf8]{inputenc}
\usepackage{amsmath}
\usepackage{amssymb}
\usepackage{color}
\usepackage[pdftex]{graphicx}

\usepackage{soul}

\begin{document}

\title{Systematic characterization of nanoscale $h$-BN quantum sensor spots created by helium-ion microscopy}
\author{Hao Gu}
\affiliation{Department of Physics, The University of Tokyo, 7-3-1 Hongo, Bunkyo, Tokyo 113-0033, Japan}
\author{Moeta Tsukamoto}
\affiliation{Department of Physics, The University of Tokyo, 7-3-1 Hongo, Bunkyo, Tokyo 113-0033, Japan}
\author{Yuki Nakamura}
\affiliation{Department of Physics, The University of Tokyo, 7-3-1 Hongo, Bunkyo, Tokyo 113-0033, Japan}
\author{Shu Nakaharai}
\affiliation{Department of Electric and Electronic Engineering, Tokyo University of Technology, 1404-4 Katakuramachi, Hachiohji, Tokyo 192-0982, Japan}
\author{Takuya Iwasaki}
\affiliation{Research Center for Materials Nanoarchitectonics, National Institute for Materials Science, 1-1 Namiki, Tsukuba, Ibaraki 305-0044, Japan}
\author{Kenji Watanabe}
\affiliation{Research Center for Electronic and Optical Materials, National Institute for Materials Science, 1-1 Namiki, Tsukuba Ibaraki 305-0044, Japan}
\author{Takashi Taniguchi}
\affiliation{Research Center for Materials Nanoarchitectonics, National Institute for Materials Science, 1-1 Namiki, Tsukuba Ibaraki 305-0044, Japan}
\author{Shinichi Ogawa}
\affiliation{National Institute of Advanced Industrial Science and Technology, 1-1-1 Umezono, Tsukuba, Ibaraki 305-8568, Japan}
\author{Yukinori Morita}
\affiliation{National Institute of Advanced Industrial Science and Technology, 1-1-1 Umezono, Tsukuba, Ibaraki 305-8568, Japan}
\author{Kento Sasaki}
\affiliation{Department of Physics, The University of Tokyo, 7-3-1 Hongo, Bunkyo, Tokyo 113-0033, Japan}
\author{Kensuke Kobayashi}
\affiliation{Department of Physics, The University of Tokyo, 7-3-1 Hongo, Bunkyo, Tokyo 113-0033, Japan}
\affiliation{Institute for Physics of Intelligence, The University of Tokyo, 7-3-1 Hongo, Bunkyo, Tokyo 113-0033, Japan}
\affiliation{Trans-scale Quantum Science Institute, The University of Tokyo, 7-3-1 Hongo, Bunkyo, Tokyo 113-0033, Japan}

\date{\today}

\begin{abstract}

The nanosized boron vacancy ($V_\mathrm{B}^-$) defect spot in hexagonal boron nitride ($h$-BN) is promising for a local magnetic field quantum sensor. 
One of its advantages is that a helium-ion microscope can make a spot at any location in an $h$-BN flake with nanometer accuracy. 
In this study, we investigate the properties of the created nanosized $V_\mathrm{B}^-$ defect spots by systematically varying three conditions: the helium-ion dose, the thickness of the $h$-BN flakes, and the substrate on which the $h$-BN flakes are attached. 
The physical background of the results obtained is successfully interpreted using Monte Carlo calculations. 
From the findings obtained here, a guideline for their optimal creation conditions is obtained to maximize its performance as a quantum sensor concerning sensitivity and localization.
\end{abstract}

\maketitle

\section{Introduction}
Investigating the magnetism arising from the cooperative
behavior of microscopic spins has been a central
topic in solid-state physics~\cite{White2007_book_quantum_state_of_magnetism}.
Many magnetic materials with various magnetic orders are known, such as ferromagnets, antiferromagnets, frustrated systems, and so on~\cite{CullityIMM2011}.
In addition to conventional bulk materials, magnetic domains, nanomagnets, atomically thin van der Waals magnets, and their application to spintronics devices tell us how diverse magnetic properties manifest themselves~\cite{CoeyHandbook2021,BurchNature2018}.
Thus, versatile methods and innovations are necessary for their experimental study.
In particular, magnetic force microscopy  and magneto-optic Kerr effect microscopy are representative methods to directly observe magnetism, which have been powerful tools for many years~\cite{KazakovaJAP2019,QiuRSI2000}.

Quantum sensors based on the nitrogen-vacancy (N-$V$) centers in diamond can detect local magnetic fields using the optically detected magnetic resonance (ODMR) technique~\cite{DegenRMP2017,Casola2018_NatRev_NV_condema}.
Either N-$V$ center ensembles or a single N-$V$ center in a scanning probe can be used to image the stray field from the target material quantitatively. 
The time-dependent magnetic responses and fluctuation in the target materials can also be measured by devising quantum control of quantum sensors.
Since the first proposals of the NV-center-based magnetometry in 2008~\cite{TaylorNaturePhys2008,BalasubramanianNature2008,DegenAPL2008}, the method has been successfully applied to condensed matter physics. 
For example, it has been used to quantitatively observe the stray fields from magnetic domain walls in antiferromagnets~\cite{HedrichNaturePhys2021,WornlePRB2021} and superconducting vortices~\cite{schlusselPRApp2018,LillieNL2020,Nishimura_APL2023_vortex}, which has been challenging to do with other existing methods.

\begin{figure}[htbp]    
    \begin{center}
    \includegraphics[width=6.5cm]{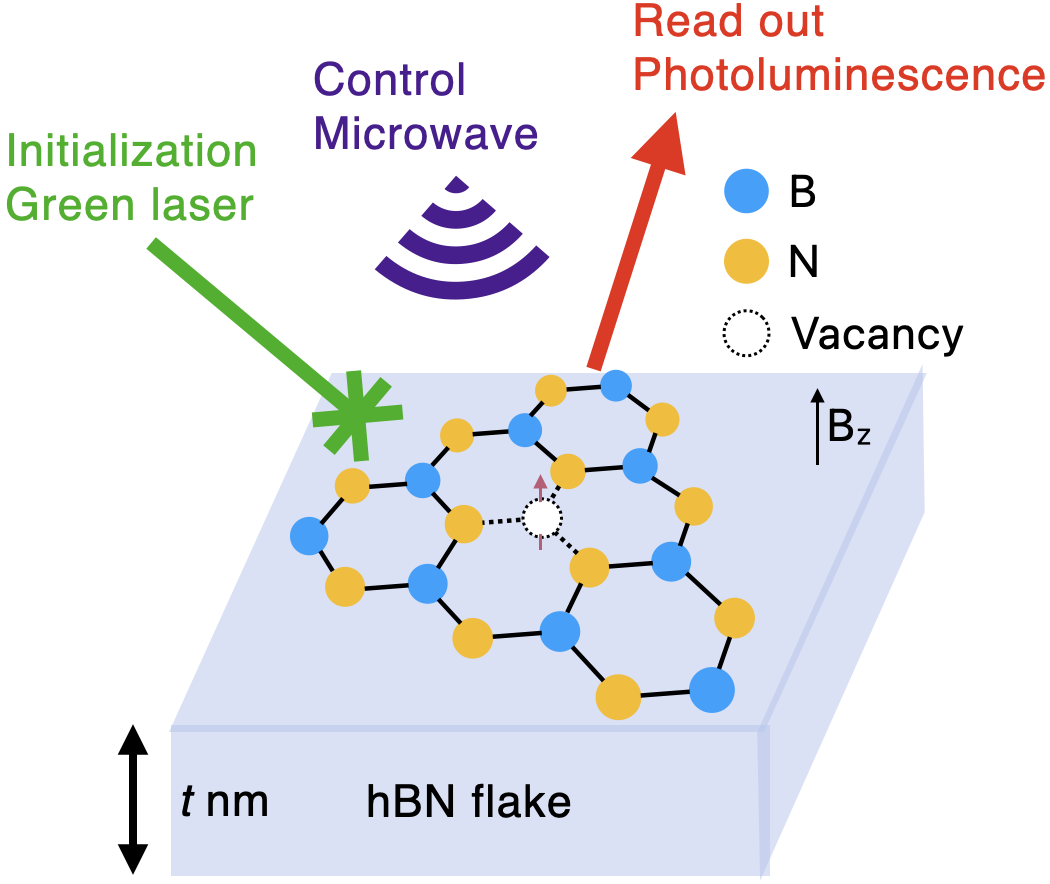}
    \caption{
    Schematic of magnetic field imaging using a $V_\mathrm{B}^-$ spot in an $h$-BN flake as thin as $t$~nm with the ODMR technique.
    \label{intro_fig}}
    \end{center}
\end{figure}

Boron vacancy ($V_\mathrm{B}^-$) defects in hexagonal boron nitride ($h$-BN), shown in Fig.~\ref{intro_fig}, were recently demonstrated to work as quantum sensors~\cite{GottschollNatMat2020,GottschollSciAdv2021, Gottscholl2021_NatComm}.
Because $h$-BN is a van der Waals material, very thinly cleaved $h$-BN flakes can easily adhere to a magnetic material to be measured.
Thus, $V_\mathrm{B}^-$ defects hosted within a flake as thin as tens of nanometers can sense magnetic fields that remain steep and strong near the sample on  nanometer order. 
Additionally, the thickness of the flake can be precisely measured, allowing us to determine the stand-off distance. 
This parameter is crucial for accurately reconstructing the magnetization and current in the target material~\cite{Broadway_PRAppl2020_reconstruction}.
For these reasons, $V_\mathrm{B}^-$ sensors are expected to be applied to detect minute magnetic orders.
Several reports on the imaging of van der Waals magnets using $V_\mathrm{B}^-$ defects created inside an $h$-BN flake or its surface have already been reported~\cite{Huang2022NatCom, HealeyNatPhys2022, Kumar2022PRAppl, Zhou_SciAdv2024_YIG_hBN}. 

To maximize the potential of the $V_\mathrm{B}^-$ sensors, we should systematically investigate both the defect creation methods and the sensing configurations.
$V_\mathrm{B}^-$ defect spots ($V_\mathrm{B}^-$ spots) are created in $h$-BN crystals by neutron~\cite{GottschollNatMat2020} or ion irradiation~\cite{Guo2022_make_vb,GuAEPX2023}.
It has been reported that the sensor properties, such as photoluminescence (PL) intensity, relaxation time, and strain, depend on irradiation conditions, including ion species, dose, and acceleration~\cite{Guo2022_make_vb}.
Therefore, investigating the damage caused by ion irradiation when creating $V_\mathrm{B}^-$ sensors is essential.
Regarding the sensing configuration, the $h$-BN flake thickness and substrate surface to which the $h$-BN flakes are attached are vital.
The flake thinner than 100~nm is conventionally used but if the flake becomes too thin, the total amount of $V_\mathrm{B}^-$ defects is reduced unfavorably.
The effect of the substrate surface is also critical, as it has been demonstrated to increase sensitivity significantly by increasing the signal intensity of $V_\mathrm{B}^-$ defects on the gold (Au) substrate film~\cite{Gao_NanoLett2021_plasmon}.
Also, in Ref.~\cite{Sarkar_NanoLett2023_HIM}, the effect of the SiO$_2$ substrate surface on the PL spectrum of $V_\mathrm{B}^-$ centers created by He ion irradiation was observed.

In addition, we emphasize the importance of the size of the created $V_\mathrm{B}^-$ spot, which directly affects the locality of the detected magnetic field.
The locality in the magnetic field detection is essential to capture changes in the magnetic field that become steeper as the spot gets closer to the target.
When $V_\mathrm{B}^-$ sensors are uniformly created in $h$-BN flakes, the measurement spot size is typically as large as the optical spot, whose size is similar to the PL wavelength (submicrometers).
To overcome this issue, Sasaki \textit{et al.}~\cite{Sasaki_APL2023_HIM} limited the actual defect spot size by ion irradiation to nanosize ($l = 25$--200~nm square), using a helium ion microscope (HIM), leading to a high locality of magnetic field detection.
This method is also advantageous as HIM can create a sensor at a designed position with nanometer accuracy.
Although some works~\cite{Liang_AOM2022_HIM, Sasaki_APL2023_HIM, Sarkar_NanoLett2023_HIM} have investigated $V_\mathrm{B}^-$ creation using a HIM, a systematic investigation of spin properties and substrate effects at practical $h$-BN flake thicknesses has not been shown.

In this study, we show the dependence of sensor properties on helium ion doses in the nanosized $V_\mathrm{B}^-$ defect creation by HIM, following our previous work~\cite{Sasaki_APL2023_HIM}.
For doses over three orders from $10^{14}$~cm$^{-2}$ to $10^{17}$~cm$^{-2}$, we systematically characterize the sensor properties of sensitivity, intensity, contrast, strain, and spin relaxation time.
Additionally, we investigate different $h$-BN flake thicknesses and substrates and observe substrate-dependent sensor properties.
We find that the static magnetic field sensitivity is best at the dose of $10^{16}$~cm$^{-2}$ for a 47~nm thick $h$-BN flake on an Au film.
We compare the experimental results with Monte Carlo simulations [Stopping and Range of Ions in Matter (SRIM)]~\cite{Ziegler2010_SRIM_first} calculating defect formation and discuss the effect of ion backscattering from the substrate.
The obtained findings provide guidelines for arranging $V_\mathrm{B}^-$ sensors using HIM.

This paper is organized as follows. 
We describe the experimental setup in Sec.~\ref{Methods_section} and explain the underlying physics for the characterization of $V_\mathrm{B}^-$ in Sec.~\ref{ODMR_mech}.
The experimental results are presented in Sec.~\ref{Results_section}.
Sections~\ref{ODMR_section} and ~\ref{Distortion_section} discuss the sensor properties obtained from ODMR spectra.
Section~\ref{T1_section} reports the spin relaxation time.
Section~\ref{SRIM} discusses the size accuracy of $V_\mathrm{B}^-$ spots and substrate dependence based on SRIM.
Section~\ref{PL_section} shows the results related to the PL intensity.
Section~\ref{Conditions_section} summarizes guidelines for creating $V_\mathrm{B}^-$ defects using HIM based on the observation and simulations shown in Sec.~\ref{Results_section}, and, finally, Sec.~\ref{Discussion_section} provides the conclusion of this work.

\section{Experiments} \label{Methods_section}

We use HIM to create $V_\mathrm{B}^-$ spots by local ion irradiation of $h$-BN flakes on a substrate. 
We systematically adopt several different fabrication conditions of $h$-BN flake thickness ($t$), substrate surface (Au or SiO$_2$), and He ion dose ($d_\mathrm{He}$) as listed in Table~\ref{conditions}.
Section~\ref{flake_handle} describes the preparation of $h$-BN flakes, Sec.\ref{He_scope} describes the $V_\mathrm{B}^-$ defect creation using HIM, and Sec.~\ref{meas} explains the confocal microscope to evaluate the properties of the created $V_\mathrm{B}^-$ spots as quantum sensors.

\begin{table}[htbp]   
    \begin{tabular}{l|c}
    \hline
    {Parameters} & Conditions \\
    \hline
    Substrate film & Au, SiO$_2$\\
    $h$-BN flake thickness, $t~(\mathrm{nm})$& $9, ~47, ~256$\\ 
    He ion dose, $d_{\mathrm{He}}$ (cm$^{-2}$) & $~10^{14}$, $10^{15}$, $10^{16}$, $10^{17}$  \\
    \hline 
    \end{tabular}
    \caption{Parameter list of the $V_\mathrm{B}^-$ defect creation.}
    \label{conditions}
\end{table}

\subsection{Device fabrication with $h$-BN flakes}\label{flake_handle}

\begin{figure}[htbp]    
    \begin{center}
    \includegraphics[width=6.5cm]{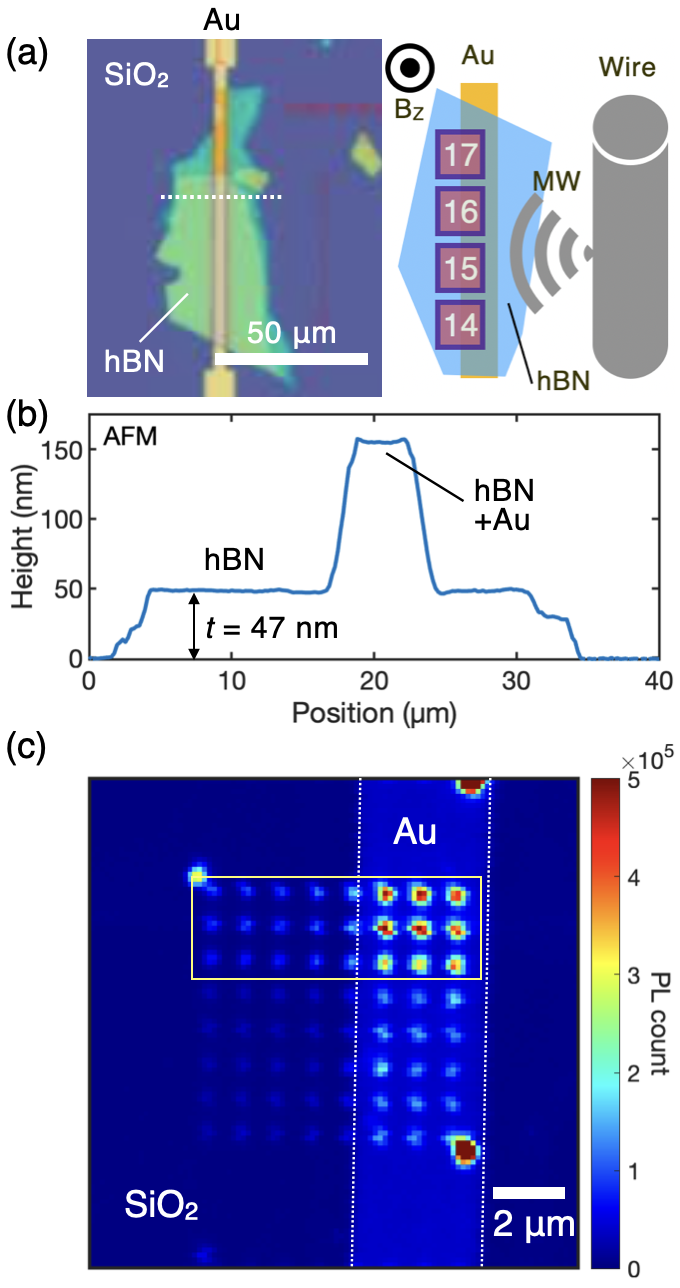}
    \caption{
    Overview of the $h$-BN device.
    (a) Left panel: Optical micrograph of the Au wire (Au film) covered by an $h$-BN flake with $t = 47$~nm. 
    Right panel: Schematic representation of the measurement configuration.
    The squares and the numbers inside them denote the irradiated regions and the exponential portion of the helium dose, namely $\log_{10} d_{\mathrm{He}}$, respectively. 
    A copper wire for applying microwaves (MW) is arranged parallel to the Au wire, as schematically shown.
    (b) AFM profile across the white dashed line in panel (a).
    (c) PL intensity mapping of the irradiation region with $d_{\mathrm{He}} = 10^{15}~\mathrm{cm^{-2}}$ of the device with $t = 256$~nm, which is obtained using a confocal microscope.
    Each spot inside the yellow frame is a 100 nm square irradiated with ions, which we evaluate in this study.
    \label{method_fig}
    }\end{center}
\end{figure}

We prepare thin $h$-BN flakes by cleaving $h$-BN bulk crystals with Scotch tape and transferring them onto a silicon substrate.
A 100 nm thick Au wire (width~4~$\mu$m) is fabricated on a silicon substrate with a 285~nm thick oxide film using photolithography, and the $h$-BN flakes are stamped on top of it using the bubble-free method~\cite{IwasakiACS2020}.
Figure~\ref{method_fig}(a) is an optical micrograph of a typical fabricated device.
The $h$-BN flake is large enough compared to the Au wire to have areas of adhesion to both Au and SiO$_2$.
An atomic force microscope (AFM) is used to measure the $h$-BN flake thickness $t$.
Figure~\ref{method_fig}(b) shows a profile corresponding to the white dashed line in Fig.~\ref{method_fig}(a). 
A few steps due to the $h$-BN flake and the Au wire (Au film) are observed.
We estimate the $h$-BN flake thickness of the device shown in Fig.~\ref{method_fig}(a) to be $t = 47$~nm. 
Similarly, we fabricate devices of $h$-BN flakes with $t = 9$~nm and $t = 256$~nm.

\subsection{$V_\mathrm{B}^-$ defect creation using HIM}\label{He_scope}

We use an Orion Plus HIM (Carl Zeiss Microscopy LLC, Peabody, MA, USA) with a helium ion beam of nominal width 0.3~nm and create $V_\mathrm{B}^-$ spots on the $h$-BN flakes on the fabricated Au wire devices.
HIM is a technique to irradiate a target object with a focused beam of helium ions for processing and imaging with high spatial resolution~\cite{Hlawacek2014_HIM_review}.
The irradiation by helium ions, which are light, is reasonable for creating small $V_\mathrm{B}^-$ spots due to less surface scattering than electron irradiation and less damage on the material structure than heavy-atom irradiation~\cite{Sarkar_NanoLett2023_HIM}.

First, we precisely determine the position of the $h$-BN flake on the Au film by observing the secondary electrons emitted from the device using HIM.
Then, the target positions at the $h$-BN flakes are irradiated with helium ions as designed in square-shaped spots of 100~nm on each side at an acceleration voltage of 30~keV.
Thus, each spot consists of many $V_\mathrm{B}^-$ defects.
The 30 keV voltage was a value used previously for the $V_\mathrm{B}^-$ defect creation with HIM~\cite{Sasaki_APL2023_HIM} and conventional ion irradiation~\cite{GuAEPX2023}.
The focused helium ion beam is discretely raster scanned with an interval of $3.2$~nm so that the average helium ion dose in the spot is $d_\mathrm{He}~(\mathrm{cm^{-2}})$.  
Each $h$-BN flake is irradiated at multiple positions in contact with Au and SiO$_2$ at $d_{\mathrm{He}}=$ 10$^{14}$, 10$^{15}$, 10$^{16}$, and 10$^{17}$~cm$^{-2}$, as shown in the right panel of Fig.~\ref{method_fig}(a) (see also Table~\ref{conditions}).
Only for the device with $t = 9$~nm, we use $d_{\mathrm{He}} = 10^{15}$, $10^{16}$, and $10^{17}$~$\mathrm{cm^{-2}}$.

\subsection{Confocal microscope system}\label{meas}

We utilize a home-built confocal microscope system~\cite{MisonouAIPAdv2020} to characterize the created $V_\mathrm{B}^-$ spots.
The PL of $V_\mathrm{B}^-$ defects occurs in the wavelength range of 750--1000~nm~\cite{GottschollNatMat2020}, and it can be detected using a bandpass filter and a single photon counting module while irradiating a green laser.
The laser wavelength is 515--532~nm for Sec.~\ref{PL_section} and 532~nm otherwise. The laser power is 0.7~mW in ODMR and spin relaxation time measurements.
It is sufficiently weaker than the typical power (on Au, 7.6~mW and on SiO$_2$, far stronger laser power is needed to saturate) at which the PL from $V_\mathrm{B}^-$ defects saturates in our confocal system.
Only in Sec.~\ref{PL_section}, the laser power is set to 3.0~mW to increase the signal intensity.

Figure~\ref{method_fig}(c) displays an example of the PL intensity mapping of a fabricated device with $t = 256$~nm and $d_{\mathrm{He}} = 10^{15}~\mathrm{cm^{-2}}$. 
Bright spots correspond to the created $V_\mathrm{B}^-$ spots.
The spots surrounded by the yellow box in Fig.~\ref{method_fig}(c) are 100 nm square size irradiated spots.
The PL intensity of the spots is more prominent on Au than on SiO$_2$.

To perform ODMR, we apply microwaves (MW) and a static magnetic field to the devices.
The MW is irradiated from a 50~$\mu$m-diameter copper wire beside the Au wire, as shown in the right panel of Fig.~\ref{method_fig}(a).
This configuration allows a strong and uniform MW irradiation to the $V_\mathrm{B}^-$ spots.
We examine spots with different doses $d_\mathrm{He}$ made on $h$-BN of different thicknesses $t$ (see Table~\ref{conditions}).
They are located along the same copper wire, as shown in the right panel of Fig.~\ref{method_fig}(a). 
Thus, the influence of MW amplitude variation on the sensitivity evaluation is minimized.
The static magnetic field is applied using a coil.
The direction of the coil's magnetic field is perpendicular to the surface of the $h$-BN flake and parallel to the quantization axis of $V_\mathrm{B}^-$, and its maximum intensity is 12~mT.
We measure spin relaxation time by pulsing MW and the laser, as described in Sec.~\ref{T1_section}.

\section{Characterization of boron vacancy defects} \label{ODMR_mech}

In this study, we measure five parameters, PL intensity ($I$), PL contrast ($C$), strain ($E$), resonance linewidth ($\Delta \nu$), and spin relaxation time ($T_1$).
This section explains their meanings and the evaluation methods.

\begin{figure}[htbp]
    \begin{center}
    \includegraphics[width=7cm]{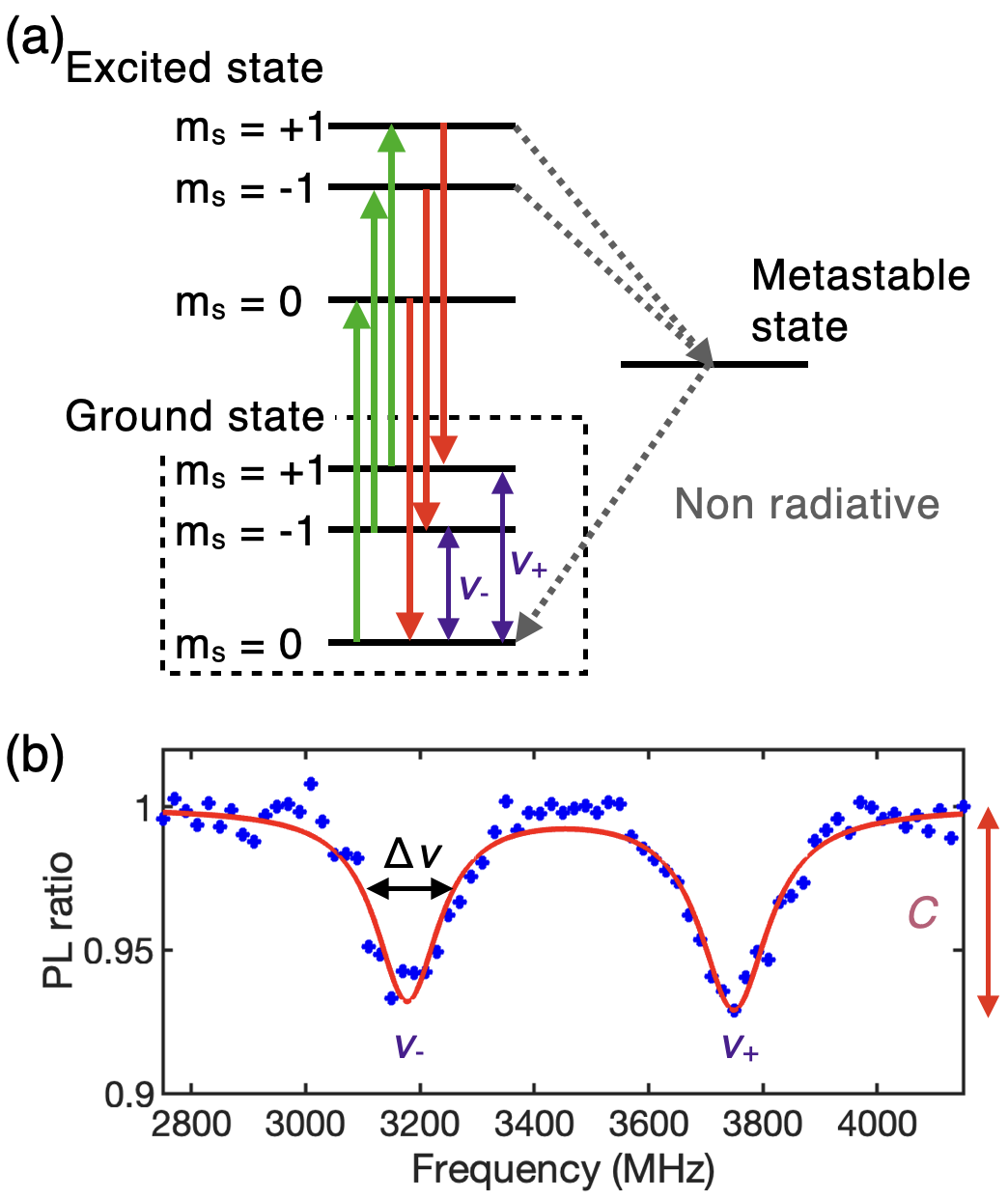}
    \caption{
    Principles of ODMR measurement.
    (a) An effective model that includes energy levels and their optical transition.
    (b) An example of the ODMR spectrum.
    The data are taken at a spot irradiated with $d_{\mathrm{He}} = 10^{16}~\mathrm{cm^{-2}}$ in the device with $t=9$~nm on the Au film. The applied magnetic field in this case is 10.2~mT.
    The blue markers denote experimental data, and the red line represents the result of the double-Lorentzian fit.
    \label{odmr_mech_fig}}
    \end{center}
\end{figure}

We determined  $C$, $E$, and $\Delta \nu$ using ODMR measurement.
We briefly outline the principle of the ODMR measurement based on the effective model of energy levels and optical transition, which includes the ground state, the excited state, and the metastable state, as illustrated in Fig.~\ref{odmr_mech_fig} (a)~\cite{robledo2011spin,gupta2016}. 
We apply this model, initially  developed for diamond N-$V$ centers, to $V_\mathrm{B}^-$ defects, assuming that they behave as an $S=1$ system, with the ground and excited states as spin triplets and the metastable state as a spin singlet.
The spin triplet is quantized in the out-of-plane direction of the $h$-BN flake, and $m_S=0$, $+1$, and $-1$ in that direction can be used to distinguish states as magnetic quantum numbers.

The sensor state transitions from the ground state to the excited state with the green laser while the magnetic quantum number is preserved [green arrows in Fig.~\ref{odmr_mech_fig}(a)]~\cite{GottschollNatMat2020}.
There are two relaxation pathways from the excited state to the ground state.
One pathway is the relaxation with red photon emission while maintaining the magnetic quantum number [red arrows in Fig.~\ref{odmr_mech_fig}(a)].
This emission constitutes the PL signal observed using a confocal microscope system.
The other pathway is through a metastable state [dashed black arrows in Fig.~\ref{odmr_mech_fig}(a)], where the magnetic quantum numbers are not conserved without  red emission.
The $m_S = \pm1$ state in the excited state selectively goes through this pathway to the $m_S = 0$ state in the ground state.
Because of this selectivity, the $m_S = \pm 1$ state exhibits a weaker PL intensity than the $m_S = 0$ state.
In addition, the sensor state is polarized or initialized to the $m_S = 0$ state with repeated excitation.

The ODMR measurement relies on the above behaviors, utilizing continuous irradiation of green light and MW.
The optical excitation initializes the $m_S = 0$ state and simultaneously yields the PL intensity.
When the MW frequency matches the electron spin resonance frequency of the ground state, a part of the $m_S = 0$ state transitions to the $m_S = \pm1$ state, leading to the PL intensity reduction.
Therefore, the intensity as a function of the MW frequency, the so-called ODMR spectrum, corresponds to an electron spin resonance spectrum.
Figure~\ref{odmr_mech_fig}(b) is an example ODMR spectrum of $V_\mathrm{B}^-$ under a magnetic field.
The vertical axis is the ratio of the PL intensity with and without MW irradiation.
Two dips correspond to the resonance between the $m_S = 0$ state and the $m_S = \pm 1$ states.
The resonance frequency and the linewidth (full width at half maximum) are $\nu_\pm$ and $\Delta\nu$, respectively.
The amount of the PL ratio change is the contrast $C$.

In quantum sensing, we estimate the magnetic field based on electron spin resonance frequencies $\nu_\pm$.
The Hamiltonian of the spin-triplet in the ground state is given by~\cite{doherty2013}
\begin{equation}
    \label{hamiltonian}
    \hat{H} = D_{\mathrm{gs}} \hat{S}_z^2 + E(\hat{S}_x^2 - \hat{S}_y^2) + \gamma_e B_z \hat{S}_z,
\end{equation}
where $\hat{S}_{j}$ is the $S=1$ operator for $j$ direction ($j=x, y, z$), $D_\mathrm{gs}$ is the zero-field splitting, $E$ is the strain, and the $\gamma_e = 28~\text{MHz/mT}$ is the electron gyromagnetic ratio. $B_z$ is the magnetic field applied along the $z$ direction, which is the out-of-plane direction of the $h$-BN flake [see right panel of Fig.~\ref{method_fig}(a)].
The second term, $E$, arises when defect symmetry is broken by a crystal strain or an electric field from charge impurities.
Thus, it can depend on irradiation damage~\cite{KleinsasserAPL2016}.
The third term is the Zeeman term, which gives rise to the magnetic field dependence of the sensor.
Here, we neglect the influence of nuclear spins near the $V_\mathrm{B}^-$ defects, which additionally split energy levels~\cite{GottschollNatMat2020,Gottscholl2021_NatComm,GottschollSciAdv2021}.
Since this does not significantly impact the discussion in our study, we do not give a detailed explanation.
By diagonalizing the Hamiltonian, the resonance frequencies are obtained as
\begin{equation}
\label{zeeman_eq}
\nu_{\pm} = D_\text{gs} \pm \sqrt{ (\gamma_e B_z)^2 + E^2 }.
\end{equation}
Therefore, if the strain $D_\text{gs}$ and $E$ are known, we can determine the magnetic field $B_z$ from the difference.

How sensitive the resonance frequency is affected by magnetic fields defines the sensitivity, as follows,
\begin{equation}
\label{zeeman_eq_grad}
\left| \frac{\partial \nu_{\pm} }{\partial B_z} \right| = \left| \frac{\gamma_e^2 B_z}{ \sqrt{(\gamma_e B_z)^2 + E^2} } \right|.
\end{equation}
This equation shows that sensitivity depends on the strain $E$.
The value decreases as the effect of magnetic field strength is sufficiently small compared to the strain ($B_z \ll E/\gamma_e$).
The strain limits the range on the low-field side where the $V_\mathrm{B}^-$ can function as a sensor; the smaller the strain, the more sensitivity the $V_\mathrm{B}^-$ can retain at lower magnetic fields.
Thus, the strain is a key parameter to characterize the sensor performance.
The strain $E$ can be estimated as half of the difference in resonance frequencies $\nu_+ - \nu_-$ near zero fields ($|B_z| \ll 0.3~\mathrm{mT}$).
The result will be discussed in Sec.~\ref{Distortion_section}.
Note that $E$ could be determined by fitting the change in resonance frequency with the magnetic field according to Eq.~(\ref{zeeman_eq}). 

The sensitivity also depends on the precision of estimating the resonance frequencies from the ODMR spectrum.
The resonance frequencies are obtained as the center frequencies of the double-Lorentzian fitted to the ODMR spectrum [the red line in Fig.~\ref{odmr_mech_fig}(b)].
The precision increases when each resonance dip is sharp, i.e., as the PL contrast $C$ increases and the linewidth $\Delta\nu$ narrows.
We deduce $C$ and $\Delta\nu$ from the lower frequency ($\nu_-$) dip.
The sensitivity is subject to the noise per unit time when obtaining the ODMR spectrum, which depends on the PL intensity $I$.
We deduce it from the PL intensity without MW, including the value obtained by PL mapping.
The static magnetic field sensitivity expression and its results will be detailed in next Sec.~\ref{ODMR_section}.
While it is beyond the scope of the present study, the sensitivity for other types of measurements is also proportional to $(C\sqrt{I})^{-1}$~\cite{GuAEPX2023}.

Note that the experimentally observed PL intensity $I$ includes $V_\mathrm{B}^-$ fluorescence $I_\text{s}$ and other background signals $I_\text{b}$.
The background contributes to the reduction of $C$.
We estimate and discuss the amount of the $V_\mathrm{B}^-$ created using HIM based on $I_\mathrm{s}$ in Sec.~\ref{PL_section}.

Finally, we explain the spin relaxation time $T_1$ as a key parameter of sensor performance.
This time $T_1$ defines the time it takes for the spin state to reach thermal equilibrium.
In the experiment, the laser is turned off once the spin is optically initialized to the $m_S = 0$ state. 
After a waiting time, $\tau$, we measure how much $m_S = 0$ states remain.
The decay of the PL intensity $I_w(\tau)$ at readout laser pulse as a function of $\tau$ behaves exponentially as
\begin{equation}
    \label{T1_eq}
    I_w(\tau) \propto \exp(-\tau/T_1) + \mathrm{offset}.
\end{equation}
The longer $T_1$ becomes, the longer the upper limit of sensing duration is, which is advantageous regarding sensitivity and frequency resolution of ac magnetic field and magnetic field noise detection~\cite{DegenRMP2017}.
The spin relaxation time is suppressed with the ion irradiation dose~\cite{Gong_NatComm2023_hBN_pulse}.
Section~\ref{T1_section} discusses $T_1$.

\section{Results and Discussions} \label{Results_section}

\subsection{Magnetic field sensitivity using ODMR} \label{ODMR_section}

We discuss the ODMR results obtained on the $V_\mathrm{B}^-$ spots on the device with a flake thickness $t = 47$~nm, which is a suitable thickness for $h$-BN flakes in magnetic imaging applications~\cite{Sasaki_APL2023_HIM}.
We estimate the shot-noise-limited static magnetic field sensitivity $\eta$ through ODMR measurements under a sufficient bias field ($B_z \gg E/\gamma_e$), using the following expression~\cite{DreauPRB2011,Barry_RevModPhys2020_NVsensitivity},
\begin{equation}
    \eta = \frac{4}{3\sqrt{3}}\frac{1}{\gamma_e}\frac{\Delta \nu}{C\sqrt{I}}.
    \label{sensitivity_eq}
\end{equation}
The increase in the contrast $C$ and PL intensity $I$ and the decrease in $\Delta \nu$ directly contribute to the increase in sensitivity.

\begin{figure}[htbp]   
    \begin{center}
    \includegraphics[width=7.5cm]{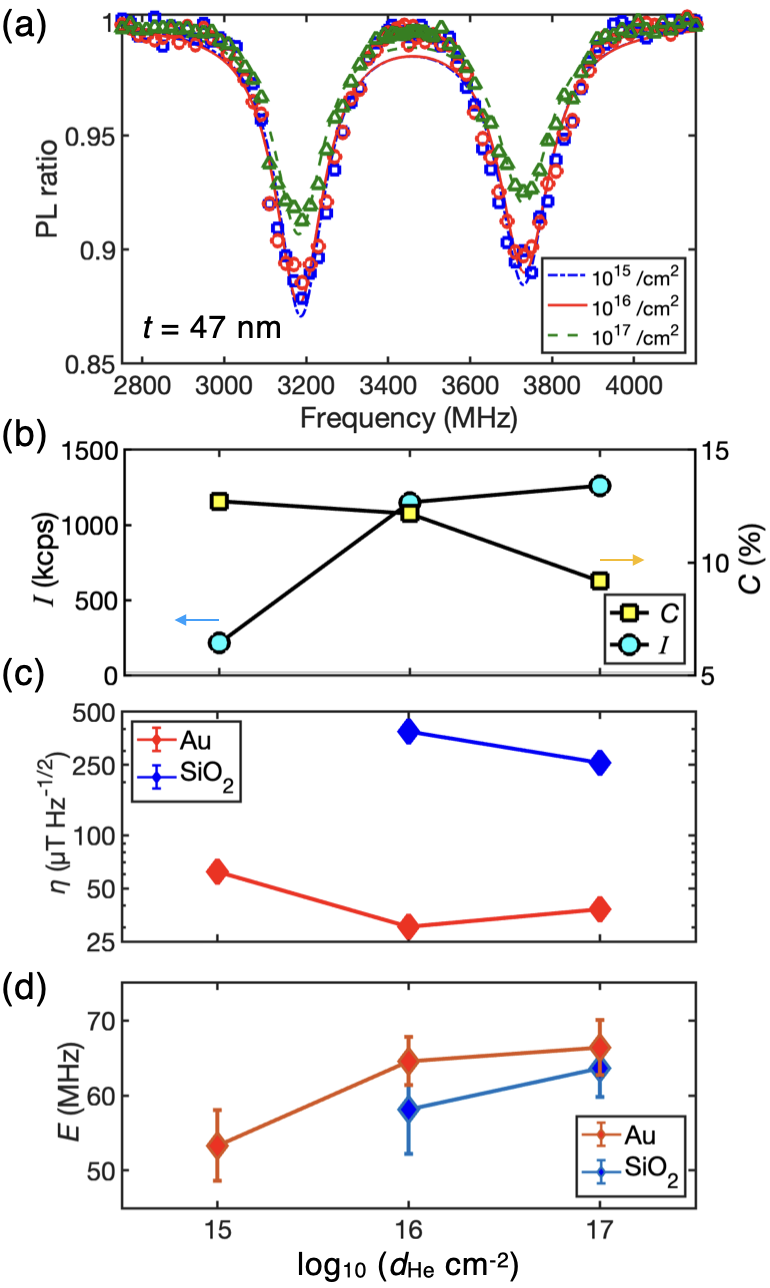}
    \caption{
    Analysis of ODMR spectrum and sensitivity for the spots with $t = 47$~nm.
    (a) ODMR spectra under $B_z = 9.7$~mT from three devices with different $d_{\mathrm{He}}$. 
    The markers denote experimental data, and the lines denote double-Lorentzian fit. 
    The blue, red, and green colors correspond to $V_\mathrm{B}^-$ spots created at $d_{\mathrm{He}} = 10^{15}$, 10$^{16}$, and 10$^{17}~\mathrm{cm^{-2}}$, respectively.
    (b) Dose-dependent PL intensity $I$ (cyan, left axis) and PL contrast $C$ (yellow, right axis).
    (c),(d) Dose-dependent (c) static magnetic field sensitivity $\eta$ and (d) strain $E$. The red and blue markers denote the spots on Au and SiO$_2$, respectively.
    \label{sensitivity}
    }\end{center}
\end{figure}

Figure~\ref{sensitivity}(a) shows the ODMR spectra for the $V_\mathrm{B}^-$ spots on Au with various doses $d_{\mathrm{He}}$.
We note that $C$ decreases as the dose increases.
We obtain $C=12.7$, 12.2, and 9.2 for $d_\mathrm{He}= 10^{15}$, $10^{16}$, and $10^{17}~\mathrm{cm^{-2}}$, respectively, as shown in Fig.~\ref{sensitivity}(b).
The decrease in $C$ with increasing $d_\mathrm{He}$ is also observed for different $t$, which is also the case for the $h$-BN flakes on SiO$_2$. 
The degradation is likely caused by suppression of spin lifetime or by increased photoluminescence from non-$V_\mathrm{B}^-$ defects due to large $d_{\mathrm{He}}$.
Amorphous defects in the $h$-BN lattice created by He ion irradiation have been reported to affect the luminescence intensity~\cite{Sarkar_NanoLett2023_HIM}, which may be related to the present observation.

In contrast to $C$, $I$ increases monotonically  with $d_\mathrm{He}$ as shown in Fig.~\ref{sensitivity}(b). 
We obtain $I = 2\times 10^2$~kcps, $1.1\times 10^3$~kcps, and $1.3\times 10^3$~kcps for $d_\mathrm{He}= 10^{15}~\mathrm{cm^{-2}}$, $10^{16}~\mathrm{cm^{-2}}$, and $10^{17}~\mathrm{cm^{-2}}$, respectively.
The estimated background $I_\mathrm{b}$ is sufficiently weak $< 20~\mathrm{kcps}$ compared to the $V_\mathrm{B}^-$ signal [see Sec.~\ref{PL_section}]. 
Therefore, the number of $V_\mathrm{B}^-$ defects increases with the amount of dose.
In this way, there is a clear trade-off between $C$ and $I$ in the dose range we investigate. 
Nevertheless, the increase of $I$ is not proportional to $d_\mathrm{He}$; the intensity is increased 5.4 times from $d_\mathrm{He}= 10^{15}$ to $10^{16}~\mathrm{cm^{-2}}$, while it is only 1.1 times from $d_\mathrm{He}= 10^{16}$ to $10^{17}~\mathrm{cm^{-2}}$.
The aforementioned amorphous defects may have prevented the formation of $V_\mathrm{B}^-$~\cite{Sarkar_NanoLett2023_HIM}.

The linewidth $\Delta \nu$ is almost insensitive to $d_\mathrm{He}$, as shown in Fig.~\ref{sensitivity}(a); only a 5 variation in $\Delta \nu$ ($136$--$143~\mathrm{MHz}$) is detected within the investigated $d_\mathrm{He}$.
This implies that the nuclear spin primarily determines $\Delta \nu$. 
Since the resonance is broadened by huge (several hundred MHz) level splitting due to nuclear spins~\cite{GottschollNatMat2020,GottschollSciAdv2021,Gottscholl2021_NatComm,GuAEPX2023}, the influence of the other factors is negligibly small.
Further investigation of the ODMR spectra for the isotope-controlled $h$-BN~\cite{Sasaki_APEX2023_isotope_hBN,Clua-ProvostPRL2023,GongNatComm2024}, where nuclear spins have less impact than conventional ones, might enable us to observe the dose dependence of linewidths.

We estimate the sensitivity $\eta$ using Eq.~(\ref{sensitivity_eq}), as shown in Fig.~\ref{sensitivity}(c). It is obtained to be 61.9~$\mathrm{\mu T/\sqrt{Hz}}$, 30.2~$\mathrm{\mu T/\sqrt{Hz}}$, and 38.2~$\mathrm{\mu T/\sqrt{Hz}}$ for $d_\mathrm{He} = 10^{15}~\mathrm{cm^{-2}}$, 10$^{16}~\mathrm{cm^{-2}}$, and $10^{17}~\mathrm{cm^{-2}}$, respectively.
In Ref.~\cite{Sasaki_APL2023_HIM}, the sensitivity of $73.6~\mathrm{\mu T/\sqrt{Hz}}$ is obtained for the spots on Au in the $h$-BN flake with $t = 66~\mathrm{nm}$ irradiated with $d_{\mathrm{He}} = 10^{15}~\mathrm{cm^{-2}}$, being consistent with the present observation (61.9~$\mathrm{\mu T/\sqrt{Hz}}$) obtained for the similar condition.
We get the best sensitivity, the minimum value of $\eta$, of $30.2~\mathrm{\mu T/\sqrt{Hz}}$ for the spots on Au with $d_\mathrm{He}=10^{16}~\mathrm{cm^{-2}}$.
The $V_\mathrm{B}^-$ density estimated by the SRIM simulation is $6.1 \times 10^{16}$ in the single spot.
The sensitivity is 2.5 times better than before~\cite{Sasaki_APL2023_HIM}.
It is an advantage of HIM that a dose as high as $10^{16}~\mathrm{cm^{-2}}$ can be realized with local ion irradiation in a reasonable time and cost.

Figure~\ref{sensitivity}(c) also shows the sensitivity of the $V_\mathrm{B}^-$ spots on SiO$_2$.
While the MW intensity is expected to differ significantly between on a metal (Au) and on an insulator (SiO$_2$), the contrast of the spots on SiO$_2$ is maintained at around $71$--$83\%$ of that on Au (not shown here). 
In contrast, $I$ on SiO$_2$ is only about $1$--$3\%$ of that on Au, as discussed later in Sec.~\ref{PL_section}.
As a result, $\eta$ is approximately an order of magnitude worse than the optimal sensitivity obtained on Au [Fig.~\ref{sensitivity}(c)].

\subsection{Strain} \label{Distortion_section}

Figure~\ref{sensitivity}(d) shows the $d_\text{He}$ dependence of the strain $E$ obtained for the spots on the device with $t = 47$~nm. 
We notice two facts.
First, increasing $d_\mathrm{He}$ leads to an increase in strain.
For example, the strains of the spots on Au are obtained as 53~MHz and 66~MHz for $d_{\mathrm{He}} = 10^{15}~\mathrm{cm^{-2}}$ and $10^{17}~\mathrm{cm^{-2}}$, respectively.
Similarly, we get $E$ on SiO$_2$ as 58~MHz and 64~MHz for $d_{\mathrm{He}} = 10^{16}~\mathrm{cm^{-2}}$ and $10^{17}~\mathrm{cm^{-2}}$, respectively.
Second, for a given $d_\mathrm{He}$, the spots on Au exhibit $3$--$10\%$ larger $E$ than those on SiO$_2$.
This observation might indicate the substrate-dependent damage to the $h$-BN flakes.
We will consider these results with simulation in Sec.~\ref{SRIM}. 

In Ref.~\cite{Guo2022_make_vb}, the dose dependence of nitrogen irradiation at an accelerating voltage of 30~keV to the 10--100 nm thick $h$-BN flakes on a silicon substrate was investigated.
They observed that strain $E$ increases from about 60 to 80~MHz as the dose increases from $10^{13}$ to $10^{15}~\mathrm{cm^{-2}}$.
In contrast, $E$ is as small as 60--65~MHz for the dose of $10^{17}~\mathrm{cm^{-2}}$ in our study.
The difference may be mainly due to the difference in the ion mass.
The nitrogen ion is about seven times heavier than the helium ion, leading to more considerable irradiation damage in the $h$-BN flakes.

The large error bars in Fig.~\ref{sensitivity}(d) are due to the double-Lorentzian fit of the  ODMR spectra near zero fields in determining strain $E$.
The double-Lorentzian shape is insufficient to reproduce the experimentally observed ODMR spectra, so the fitting precision needs to be more satisfactory.
A more appropriate analytical expression for zero-field ODMR spectra, such as discussed before~\cite{MatsuzakiJPCM2016}, will enhance the estimation precision of the sensor parameters.

\subsection{Spin relaxation time} \label{T1_section}

\begin{figure}[htbp]   
    \begin{center}
    \includegraphics[width=7.5cm]{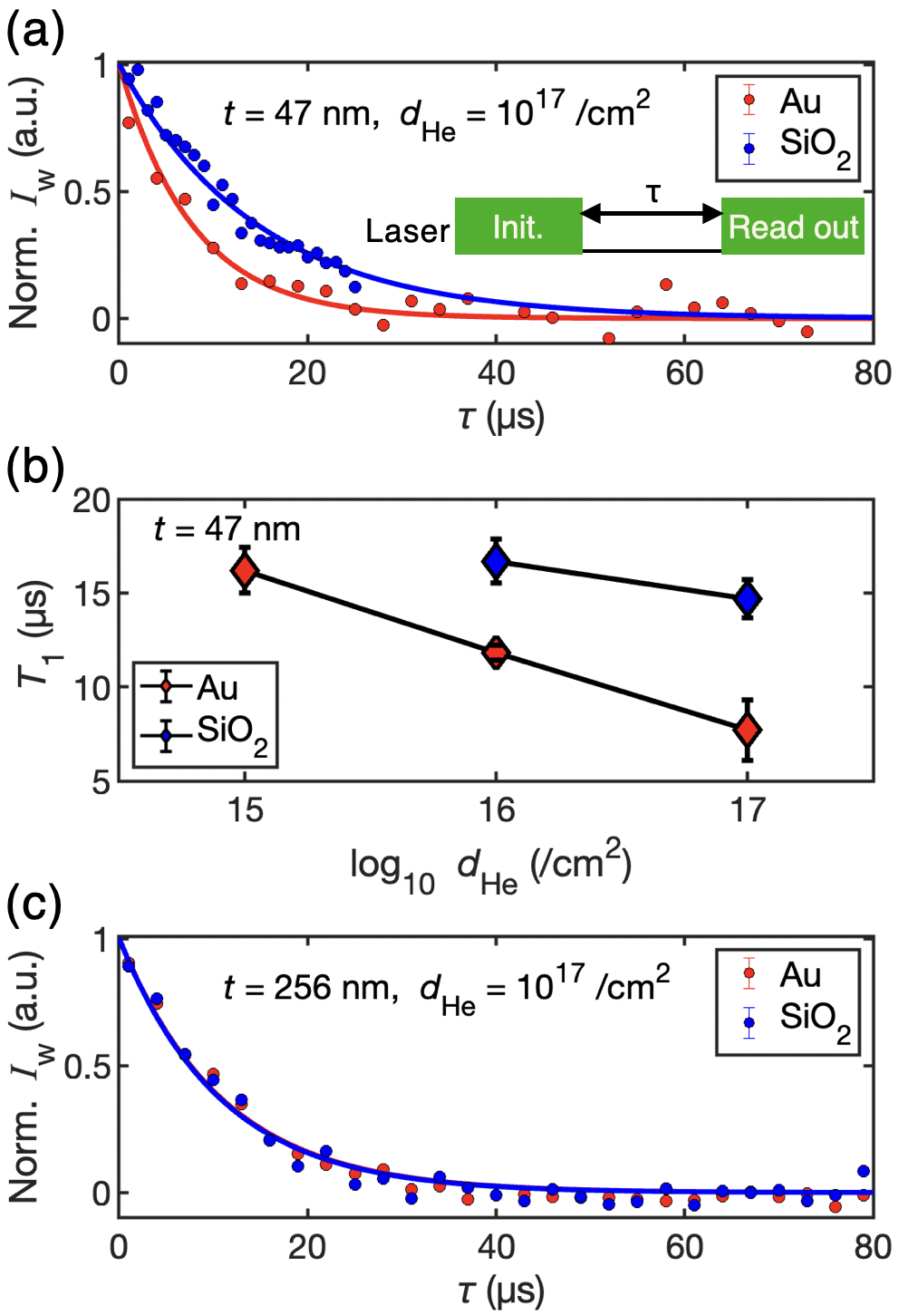}
    \caption{
    (a) Relaxation behavior of a $V_\mathrm{B}^-$ spot with $t = 47$~nm and $d_{\mathrm{He}} = 10^{17}~\mathrm{cm^{-2}}$. 
    Red and blue colors represent the results on Au and on SiO$_2$, respectively. 
    Points represent experimental data, and lines indicate exponential fits using Eq.~(\ref{T1_eq}). 
    The inset shows the measurement protocol, consisting of initialization,  waiting ($\tau$), and readout.
    (b) Spin relaxation time $T_1$ as a function of $d_{\mathrm{He}}$. 
    Red and blue markers indicate the spots on Au and SiO$_2$, respectively.
    (c) Relaxation behavior corresponding to panel (a) in the case of $t = 256$~nm. 
    \label{T1_fig}
    }\end{center}
\end{figure}

We measure the spin relaxation time $T_1$ using the pulse protocol depicted in Fig.~\ref{T1_fig}(a) inset.
The spin state population is estimated from the PL intensity at the first 200~ns of the readout laser pulse.
Figure~\ref{T1_fig}(a) shows the results for the spots irradiated with $d_\mathrm{He} = 10^{17}~\mathrm{cm^{-2}}$ at the device with $t=47$~nm.
The vertical axis is the normalized PL intensity $I_w$ so that $I_w = 1$ at $\tau = 0$~ns and $I_w = 0$ when $\tau$ is sufficiently long.
Their behaviors are well explained by Eq.~(\ref{T1_eq}).
Remarkably, the PL intensity decays faster in the spots on Au (red, $T_1 = 7.7~\mathrm{\mu s}$) than those on SiO$_2$ (blue, $T_1 =14.7~\mathrm{\mu s}$).
These values are typical for $T_1$ of $V_\mathrm{B}^-$ centers \cite{Gao_ACSPHoto2023_Gd_hBN, Huang2022NatCom, Gao_NanoLett2021_plasmon}.

We also investigate the $d_{\mathrm{He}}$ dependence of $T_1$, as presented in Fig.~\ref{T1_fig}(b).
We notice two trends.
First, $T_1$ becomes smaller for larger $d_\mathrm{He}$. 
This observation agrees with previous studies~\cite{Guo2022_make_vb, Gong_NatComm2023_hBN_pulse}, which discussed that the $T_1$ degradation is caused by lattice damage during irradiation.
For the spots on Au, changing $d_\mathrm{He}$ from $10^{15}~\mathrm{cm^{-2}}$ to $10^{16}~\mathrm{cm^{-2}}$ results in a 27$\%$ degradation, and changing from $10^{16}$ to $10^{17}~\mathrm{cm^{-2}}$ leads to a 35$\%$ degradation. 
For the spots on SiO$_2$, changing $d_\mathrm{He}$ from $10^{16}~\mathrm{cm^{-2}}$ to $10^{17}~\mathrm{cm^{-2}}$ results in an 11$\%$ degradation.
Clearly, the degradation is more pronounced for the spots on Au than on SiO$_2$.
Second, for a given $d_\mathrm{He}$, the spots on SiO$_2$ have a 1.4--1.9 times longer $T_1$ than those on Au.

Figure~\ref{T1_fig}(c) shows the spin relaxation of the spots on the device with a thicker $h$-BN of $t = 256$~nm with $d_\mathrm{He} = 10^{17}~\mathrm{cm^{-2}}$. 
In contrast to the case with $t = 47$~nm shown in Fig.~\ref{T1_fig}(a), the decays for the spots on Au and on SiO$_2$ are almost identical.

The above results suggest that irradiation damage depends on the substrate and $h$-BN thickness.
This insight is further investigated by the Monte Carlo simulations next.

\subsection{Monte Carlo simulations}\label{SRIM}

We run a Monte Carlo simulation package, Stopping and Range of Ions in Matter~\cite{Ziegler2010_SRIM_first} (SRIM), where the collision events and the resultant vacancy distribution created by ion irradiation are calculated.
SRIM treats atomic collisions as classical two-body ones, including the atomic interactions and the cascade effect where one scattered atom scatters another.
Since the ion irradiation spot size of HIM is extremely small, the spreading effect of ions randomly colliding in the target material should be carefully treated in estimating the actual $V_\mathrm{B}^-$ spot size.

\begin{figure}[htbp]   
    \begin{center}
    \includegraphics[width=7cm]{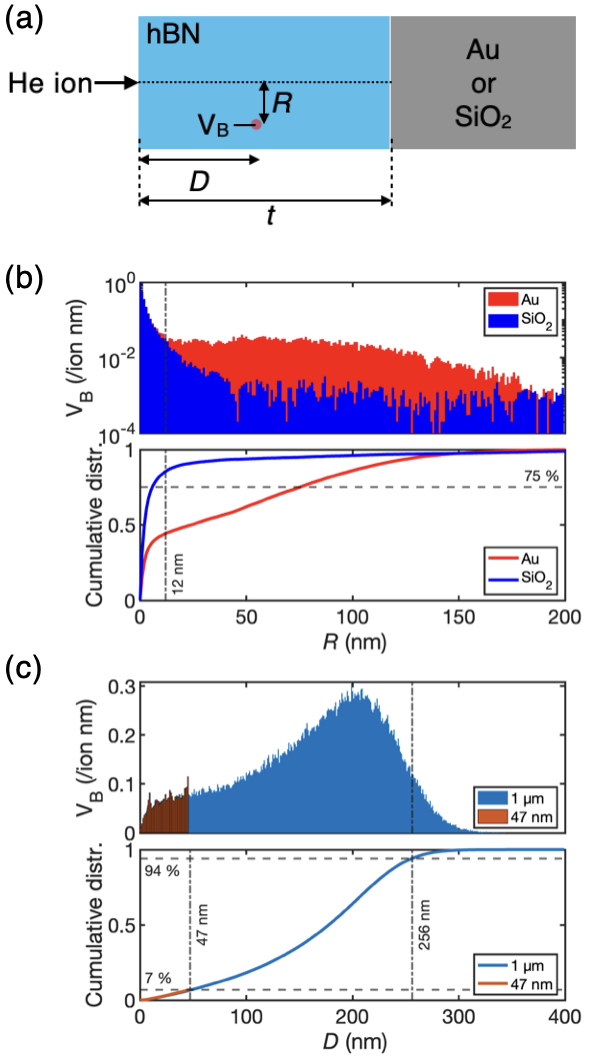}
    \caption{
    (a) Configuration of ion irradiation in SRIM.
    Ions enter the $h$-BN flake perpendicularly from the incident position.
    The distance from the incident axis and the depth from the surface are indicated by $R$ and $D$, respectively.
    The flake is attached to a sufficiently thick Au or SiO$_2$ film. 
    (b) Top panel: Histogram of the $V_\mathrm{B}^-$ defects created per unit length for $R$ of the device with $t = 47$~nm.
    Bottom panel: Normalized cumulative distribution of $V_\mathrm{B}^-$ defects.
    The horizontal dashed line indicates 75\%. 
    (c) Top panel: Histogram of the $V_\mathrm{B}^-$ defects per unit length for $D$ of the device with thick $h$-BN flake ($t = 1~\mathrm{\mu m}$).
    The blue and red colors indicate the cases of thick ($1~\mathrm{\mu m}$) and thin ($47~\mathrm{nm}$) $h$-BN flakes on SiO$_2$, respectively. 
    The vertical dotted lines represent the two thicknesses (47 and 256~nm) used in our experiment.
    Bottom panel: Normalized cumulative distribution of $V_\mathrm{B}^-$ defects. 
    The $h$-BN flake thickness does not affect the distribution of $V_\mathrm{B}^-$ defects in the  SiO$_2$ case. 
    \label{SRIM_fig}
    }\end{center}
\end{figure}

Figure~\ref{SRIM_fig}(a) shows the simulation configuration, where helium ions are directed perpendicular to the $h$-BN flake from left to right and enter it perpendicularly from the incident position.
There, $R$ and $D$ indicate the distance from the incident axis and the depth from the surface, respectively.
The flake is attached to a sufficiently thick Au or SiO$_2$ film. 
The backscattering of ions at the substrate film plays an important role in the defect creation.
We set the acceleration voltage to 30~keV and the $h$-BN, SiO$_2$, and Au densities at 2.3~g/cm$^3$~\cite{Sichel_PRB1976_hBN_structure}, 2.1~g/cm$^3$~\cite{Sze_book}, and 19.3~g/cm$^3$.
We consider $V_\mathrm{B}^-$ positions to be those of the removed boron atoms. 
We analyze the results of irradiating each device with a total of 10$^4$ helium ions.

The top panel of Fig.~\ref{SRIM_fig}(b) shows the histogram of the density of the created $V_\mathrm{B}^-$ defects per unit length as a function of $R$ for the device with $t = 47$~nm.
In the range $R < 12$~nm [indicated by the vertical dot-dash line in Fig.~\ref{SRIM_fig}(b)], the $V_\mathrm{B}^-$ defect density distribution is nearly the same for the $h$-BN flakes on Au and SiO$_2$. 
In contrast, for $R > 12$~nm, more $V_\mathrm{B}^-$ defects tend to be created for the flake on Au than on SiO$_2$.
In total, the $V_\mathrm{B}^-$ defects are created 2.0 times more on the former than on the latter, which means that the total damage is higher for the flake on Au.
Au has a higher density than SiO$_2$, so backscattering is more significant, leading to increased irradiation damage.
In Secs.~\ref{Distortion_section} and ~\ref{T1_section}, we discussed larger $E$ and shorter $T_1$ for devices on Au than on SiO$_2$.
This is consistent with our simulation that lattice damage depends on the substrate film.

The bottom panel of Fig.~\ref{SRIM_fig}(b) shows the normalized cumulative distribution as a function of $R$.
For the $h$-BN ($t = 47$~nm) on Au, three-fourths (75\%) of all $V_\mathrm{B}^-$ defects are created scattered over an area up to $R~\sim 76$~nm, as indicated by the horizontal dashed line in the figure.
Such a spreading of the $V_\mathrm{B}^-$ defects degrades the localization of the sensor.
In contrast, on SiO$_2$, the same amount of $V_\mathrm{B}^-$ defects is concentrated only in an $R<6$~nm area.
Using the SiO$_2$ substrate with less backscattering would help create a $V_\mathrm{B}^-$ spot as small as a few~nm.

Figure~\ref{SRIM_fig}(c) shows the depth ($D$) dependence  of the defect creation for a thick $h$-BN flake ($t = 1~\mathrm{\mu m}$).
Almost all of $V_\mathrm{B}^-$ defects are created shallower than $D = 256$~nm in a thick $h$-BN flake, as indicated by a vertical dashed line in the bottom panel of Fig.~\ref{SRIM_fig}(c).
This means that the backscattering from the substrate (Au or SiO$_2$) is almost negligible in an $h$-BN flake thicker than this.
The difference in the defect creation between Au and SiO$_2$ in a 256~nm thick $h$-BN flake is only 1\% (data not shown). 
It is also interesting to focus on a small $D$ region.
The normalized cumulative distribution from the surface is only 7\% at depth $D = 47$~nm.
Thus, for the thin $h$-BN flake with $t = 47$~nm, most $V_\mathrm{B}^-$ defects are produced by ions that have undergone backscattering.

The above observations can explain several experimental findings discussed in Secs.~\ref{Distortion_section} and ~\ref{T1_section}.
Au tends to backscatter ions more than SiO$_2$, so a thin $h$-BN flake on Au is more subject to backscattering damage than one on SiO$_2$. 
This agrees with a larger strain $E$ for $t= 47$~nm in the Au case than in the SiO$_2$ case, shown in Fig.~\ref{sensitivity}(d) for a given $d_\mathrm{He}$.
We can also claim that the backscattering effect is responsible for the shorter spin relaxation time $T_1$ of the $t= 47$~nm flake on Au than on SiO$_2$, as shown in Fig.~\ref{T1_fig}(a).
In sharp contrast, $T_1$ is almost the same on Au and SiO$_2$ in a thick $h$-BN flake ($t=256$~nm) [see Fig.~\ref{T1_fig}(c)], which is concordant with the calculation that the fraction of $V_\mathrm{B}^-$ defects created by backscattering is reduced for thicker flakes.
Thus, the SRIM results nicely illustrate the experimental results.

\subsection{Photoluminescence} \label{PL_section}

We evaluate the PL intensity at the irradiated spot for various conditions.
For this investigation only, we set the laser power to 3.0~mW, our maximum available power, to enhance the signal.
Note that we do not treat the result of the spots of the device with $t=47$~nm and $d_\mathrm{He} = 10^{14}~\mathrm{cm^{-2}}$ and of the device with $t=9$~nm on SiO$_2$, as they do not provide a sufficient luminescence signal from the analysis.
As mentioned in Sec.~\ref{ODMR_mech}, the overall PL intensity $I$ includes background fluorescence other than $V_\mathrm{B}^-$.
We separate the spot luminescence $I_\text{s}$ from the background $I_\text{b}$ to focus only on the increase in luminescence due to ion irradiation.

\begin{figure}[htbp]
    \begin{center}
    \includegraphics[width=7cm]{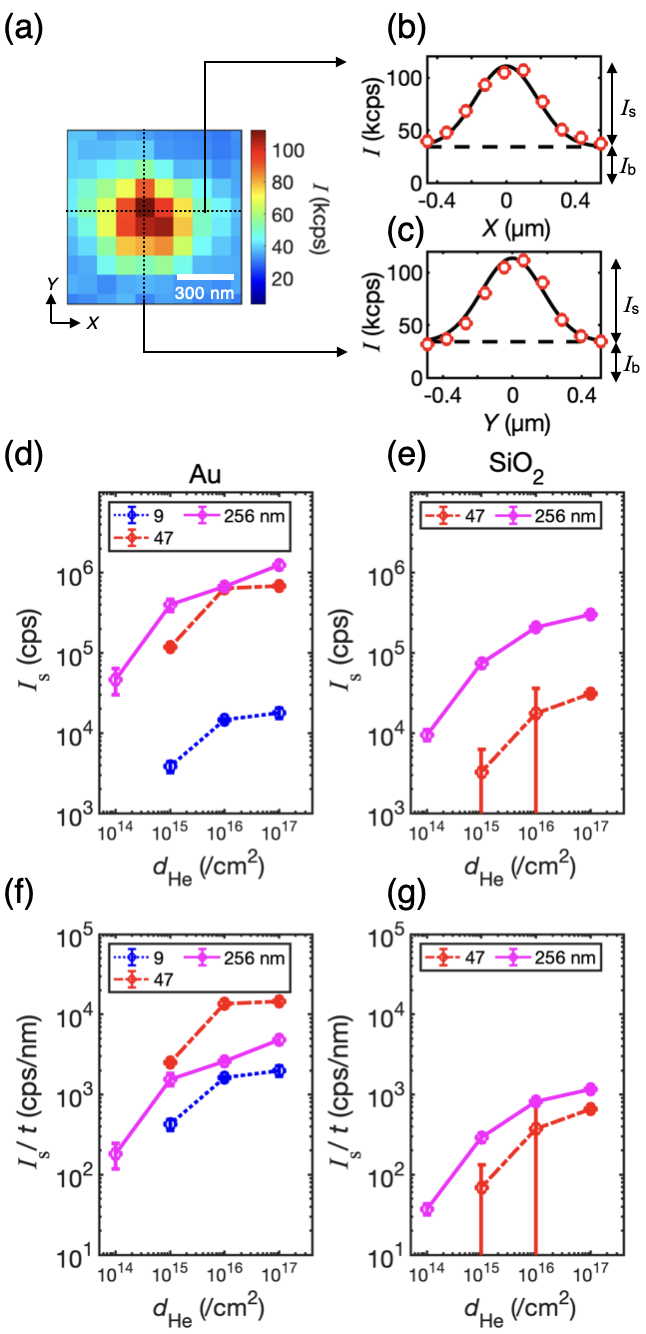}
    \caption{
    PL intensity without background $I_\mathrm{s}$.
    (a) Typical PL mapping as an $XY$ plane at a $V_\mathrm{B}^-$ spot on the $h$-BN flake with $t=256$~nm and $d_\text{He} = 10^{15}$~cm$^{-2}$. 
    (b),(c) Cross sections along the (b) horizontal ($X$) and (c) vertical ($Y$) directions across the peak. 
    The markers represent experimental data, and the lines represent the two-dimensional Gaussian fitting with an offset.
     (d),(e) The $d_{\mathrm{He}}$ dependence of $I_\mathrm{s}$ on (d) Au and (e) SiO$_2$. 
    The dotted blue, dashed red, and solid magenta lines are the data of the devices with $t = 9$, 47, and 256~nm, respectively.
    (f),(g) PL intensity per unit thickness $I_\mathrm{s}/t$ on (f) Au and (g) SiO$_2$.
    \label{PL_spotsize}
    }\end{center}
\end{figure}

We separate $I_\mathrm{s}$ from $I_\text{b}$ in the following way.
Figure~\ref{PL_spotsize}(a) shows a typical PL mapping as an $XY$ plane at a $V_\mathrm{B}^-$ spot on the $h$-BN flake. 
The spot shape can be fitted using a two-dimensional Gaussian distribution with a peak contribution of $I_\mathrm{peak}$, including an offset of $I_\mathrm{b}$.
Figures~\ref{PL_spotsize}(b) and (c) show the cross sections of the PL intensity across the peak along the $X$ and $Y$ axes, respectively. 
The markers represent the experimental data, and the lines represent the fitting result.
The dashed line corresponds to $I_\mathrm{b}$.
Then, we find $I_\mathrm{s}$ as $I_\mathrm{peak} - I_\mathrm{b}$.

Figures~\ref{PL_spotsize}(d) and (e) show the dose dependence of $I_\mathrm{s}$ for the spots on Au and SiO$_2$, respectively.
$I_\mathrm{s}$ shows a monotonic increase in $d_{\mathrm{He}}$ for all devices with $t = 9$, 47, and 256~nm.
The  rise in $I_\mathrm{s}$ becomes smaller for higher $d_{\mathrm{He}}$.
It corresponds to a decrease in $V_\mathrm{B}^-$ defect creation efficiency at high doses and agrees with the discussion given for Fig.~\ref{sensitivity}(b) in Sec.~\ref{ODMR_section}. 
It is known that amorphization of substrates and $h$-BN crystals occurs at helium doses of $10^{17}$~cm$^{-3}$ with an acceleration voltage of 30~keV~\cite{Livengood2009_SRIM_HIM_bubble, Fox2012_HIM_bubble}, and such defects may prevent $V_\mathrm{B}^-$ defect creation.

Next, we compare $I_s$ of the spots on Au and on SiO$_2$ in Figs.~\ref{PL_spotsize}(d) and (e), respectively.
For the spots with $t = 47~$nm and $d_{\mathrm{He}}=10^{17}~\mathrm{cm^{-2}}$, $I_s$ is about 22 times larger on Au than on SiO$_2$. 
The enhancement is peculiar because the simulation results in Sec.~\ref{SRIM} naively predict that the $V_\mathrm{B}^-$ defect creation efficiency on Au is about two times larger than on SiO$_2$.
This marked enhancement is due to the luminescence enhancement on Au.
In a previous study~\cite{Gao_NanoLett2021_plasmon}, for an $h$-BN flake with $t\sim50$~nm, the luminescence on Au is 10--15 times stronger than on SiO$_2$.
Thus, it is reasonable that $I_s$  increases about 20 times more on Au than on SiO$_2$. 

In contrast, the $I_s$ is only 2--3 times larger on Au than on SiO$_2$ for the device with $t = 256~$nm.
This agrees with the fact that the luminescence enhancement on Au is suppressed at an $h$-BN as thick as 200~nm~\cite{Gao_NanoLett2021_plasmon}.
Also, as discussed in Sec.~\ref{SRIM}, a thick flake has no significant enhancement due to backscattering from the substrate in the $V_\mathrm{B}^-$ defect creation.

Finally, in Figs.~\ref{PL_spotsize}(f) and (g), we compare $I_\mathrm{s}$ per unit flake thickness, i.e., $I_\mathrm{s}/t$.
In the devices on Au, the value is maximized for $t = 47~$nm, being 1.6--5.4 times higher than for $t = 256$~nm and 5.9--8.4 times higher than for $t = 9$~nm. 
This result is attributed to the composite effects of the backscattering [Sec.~\ref{SRIM}] and the $t$-dependent luminescence enhancement on Au~\cite{Gao_NanoLett2021_plasmon}.
In the devices on SiO$_2$, in contrast, $I_\mathrm{s}/t$ is bigger for $t = 256$~nm than for $t = 47$~nm. 
This is consistent with the fact that the depth at which most boron vacancies are created is $D = 150$--250~nm for an acceleration voltage of 30~keV, as shown in the top panel of Fig.~\ref{SRIM_fig}(c).
When the $h$-BN is very thin on SiO$_2$, the He ion goes through without defect creation.

\subsection{Optimum conditions} \label{Conditions_section}

We now summarize the results obtained so far and provide guidelines for creating $V_\mathrm{B}^-$ spots using HIM. 

First, we discuss the choice of the substrate. 
The advantage of selecting Au is that, even at a low $d_{\mathrm{He}}$, the effective dose increases due to significant backscattering for an $h$-BN flake with a thickness well below 256~nm. 
Additionally, Au significantly enhances the PL intensity, which is beneficial for high magnetic-field sensitivity.
The optimum dose that maximizes sensitivity in the conditions investigated in this study is  about $10^{16}~\mathrm{cm^{-2}}$ (Table  \ref{table_best_conditions}).
The optimal condition could be further investigated with different spot sizes and acceleration voltages.

\begin{table}[htbp]   
    \begin{tabular}{lr}
    \hline   
    \hline
    {Parameters} &{     }\\
    \hline
    Substrate film & Au\\
    He ion dose, $d_{\mathrm{He}}$ (cm$^{-2}$) & $10^{16}$  \\
    Acceleration voltage of He beam (keV) & $30$\\
    Sensitivity ($\mathrm{\mu T/\sqrt{Hz}}$)& 30.2\\
    \hline
    \hline
    \end{tabular}
    \caption{Best sensitivity condition for $V_\mathrm{B}^-$ on 47 nm $h$-BN.}
    \label{table_best_conditions}
\end{table}

On the other hand, the advantage of SiO$_2$ is that SiO$_2$ causes less backscattering than Au, and a well-localized spot can be created.
The spot size is expected to be smaller than 10~nm for devices with a thickness of 47~nm [see Fig.~\ref{SRIM_fig}(b)]. 
The optimum dose for this purpose is $10^{17}~\mathrm{cm^{-2}}$, which gives a sensitivity of $\sim 250~\mathrm{\mu T/\sqrt{Hz}}$ [Fig.~\ref{sensitivity}(c)].
Further fine-tuning of dose and verification at higher doses may yield even better sensitivity.
In principle, high localization and sensitivity can be obtained simultaneously by irradiating $h$-BN flakes on SiO$_2$ with helium ions to create $V_\mathrm{B}^-$ spots and then stamping them on Au.

Second, we comment on the choice of flake thickness $t$.
Consider the case where flakes are sufficiently thin (say, $t \ll 256$~nm) and irradiated at a given $d_{\mathrm{He}}$; as shown in Fig.~\ref{SRIM_fig}(c), in that case, the thicker the flake, the more the $V_\mathrm{B}^-$ defects are created. 
On the other hand, as the thickness increases, the spot size increases due to collision processes inside the flake and the backscattering at the substrate film (Au or SiO$_2$). 
Fortunately, if we use an $h$-BN flake on Au, the luminescence enhancement is substantial at $t\sim 50$~nm, so there is no need to increase the thickness of the flakes any further.

\section{Conclusion}\label{Discussion_section}
To conclude, we have investigated sensor parameters of $V_\mathrm{B}^-$ spots created using HIM under various conditions, fixing only the acceleration voltage of 30~keV.
From the experimental and simulation results, we obtain the following three findings.
First, we find the optimal dose for the $h$-BN on Au to achieve high static magnetic field sensitivity.
Second, the sensor performance depends on the substrates, Au and SiO$_2$.
The ion backscattering from Au significantly affects sensor parameters such as strain $E$ and spin relaxation time $T_1$ for thin flakes. 
However, the effect of backscattering becomes negligible in a sufficiently thick flake. 
Third, based on simulations, the $V_\mathrm{B}^-$ spot is localized more on SiO$_2$ than on Au. 
These results help optimize sensing configuration using $V_\mathrm{B}^-$ spots created with HIM.
Notably, many of the discussions in the present paper apply to general cases of $V_\mathrm{B}^-$ defect creation using HIM and conventional ion irradiation. 

The original purpose of the present work is to improve the effective spatial resolution of the stray magnetic field by localizing the sensor. 
The created quantum sensors using this approach allow adjustable and rigid determination of the stand-off distance and in-plane position between the quantum sensor and the target object.
Designing appropriate patterns for $V_\mathrm{B}^-$ spots is expected to create new probes for studying condensed matter-physics using nanosized quantum sensor spots, such as investigating microscopic spatial correlations in magnetic materials~\cite{Rovny_Science2022_covariance}.
As an application, arranging $V_\mathrm{B}^-$ spots in an array and employing a high-performance camera could enable simultaneous high-precision magnetic field imaging at multiple spots~\cite{Sasaki_APL2023_HIM}. 
Moreover, by refining the analysis methods, there is a possibility of independently examining signals from spots that approach or surpass the optical resolution.

\begin{acknowledgments}

We thank Tomohiko Iijima (AIST) for the usage of AIST SCR HIM for the helium-ion irradiations, Toshihiko Kanayama (AIST) for helpful discussions since the introduction of HIM at AIST in 2009, and Kohei M. Itoh (Keio University) for letting us use the confocal microscope system. This work was partially supported by JST, CREST Grant No. JPMJCR23I2, Japan; Grants-in-Aid for Scientific Research (Grants No. JP24KJ0692, No. JP24KJ0880, No. JP23K25800, No. JP22K03524, No. JP22KJ1059, No. JP19H00656 and No. JP19H05826); ``Advanced Research Infrastructure for Materials and Nanotechnology in Japan (ARIM)'' (Proposal No. JPMXP1222UT1131) of the Ministry of Education, Culture, Sports, Science and Technology of Japan (MEXT); the Mitsubishi Foundation (Grant No. 202310021); Kondo Memorial Foundation; JSR Corporation; Daikin Industries, Ltd.; and the Cooperative Research Project of RIEC, Tohoku University. K.W. and T.T. acknowledge support from the JSPS KAKENHI (Grants No. 21H05233 and No. 23H02052) and World Premier International Research Center Initiative (WPI), MEXT, Japan. H.G., Y.N., and M.T. acknowledge financial support from FoPM, WINGS Program, The University of Tokyo, and JSPS Young Researcher Fellowship. H.G. acknowledge financial support from JSR Fellowship, The University of Tokyo.

\end{acknowledgments}

\end{document}